\documentclass[11pt,twoside,a4paper]{article}
\usepackage{amsmath}
\usepackage{amssymb}
\usepackage{amsthm}
\usepackage{fullpage}
\usepackage{graphicx}
\usepackage{verbatim}
\makeatletter
\def\Ddots{\mathinner{\mkern1mu\raise\p@
\vbox{\kern7\p@\hbox{.}}\mkern2mu
\raise4\p@\hbox{.}\mkern2mu\raise7\p@\hbox{.}\mkern1mu}}
\makeatother
\numberwithin{equation}{section}
\title{Symmetric hyperbolic monopoles}
\author{Alexander Cockburn\\ \\
\emph{ Department of Mathematical Sciences, Durham University, Durham DH1 3LE, U.K.
}\\ \\a.h.cockburn@durham.ac.uk\\}
\date{June 2014}
\begin{document}
\begin{flushright}
\emph{DCPT-14/29}
\end{flushright}
{\let\newpage\relax\maketitle}
\begin{abstract}
Hyperbolic monopole solutions can be obtained from circle-invariant ADHM data if the curvature of hyperbolic space is suitably tuned.  Here we give explicit ADHM data corresponding to axial hyperbolic monopoles in a simple, tractable form, as well as expressions for the axial monopole fields.  The data is deformed into new 1-parameter families preserving dihedral and twisted-line symmetries.  In many cases explicit expressions are presented for their spectral curves and rational maps of both Donaldson and Jarvis type.
\end{abstract}
\section{Introduction}
Some time ago, Atiyah made the observation that hyperbolic monopoles are equivalent to circle-invariant Yang-Mills instantons if a discrete relationship exists between the curvature of hyperbolic space and the magnitude of the Higgs field at infinity \cite{A}.  A rich theory of hyperbolic monopoles has since developed, involving twistor correspondences to spectral curves and rational maps.  It was shown in \cite{BA} that monopoles satisfying Atiyah's relationship are equivalent to a discrete Nahm system called the Braam-Austin equations.  More recently, there has been interest in explicit solutions for the simplest example of Atiyah's relationship.  A large class of examples with Platonic symmetry was given in \cite{PHM}.  Explicit formulae for the spectral curves and rational maps of monopoles obtained from the JNR ansatz were obtained in \cite{BCS}, as well as more explicit solutions with Platonic symmetry and some 1-parameter families with cyclic and dihedral symmetry.

In this paper we give more monopole solutions, still within the simplest example of Atiyah's relationship, as well as their associated spectral curves and rational maps.  The new examples were found by deforming the axial monopole ADHM data while imposing judiciously chosen discrete symmetries.  This approach is only possible if one has axial monopole ADHM data in a particularly convenient form.  It was realised some time ago \cite{N} that axial monopoles are readily obtainable from the JNR ansatz, although this approach is unnatural at the ADHM level.  Here we give a different, indirect derivation of axial monopole ADHM data via the Braam-Austin construction.  The ADHM data so obtained follows a simple pattern and one can clearly see the action of the axial symmetry.  The new examples include circle-invariant ADHM data corresponding to 1-parameter families with various kinds of dihedral and twisted-line symmetry, for all values of the topological charge.

The plan of the paper is as follows. Section 2 contains a brief review of hyperbolic monopoles and their relation to Euclidean instantons.  Section 3 reviews the Braam-Austin and Manton-Sutcliffe approaches to circle-symmetric ADHM data.  In Section 4 we give the first new results, which are explicit data and Higgs fields for axial hyperbolic monopoles.  In Section 5 we deform these to give some interesting new 1-parameter families, and we make some concluding remarks in Section 6.
\section{Hyperbolic monopoles and instantons}
Hyperbolic monopoles are solutions of the Bogomolny equation
\begin{equation}\label{bog}
D\phi=*F
\end{equation}
where $F$ is the field strength of an $SU(2)$ gauge potential $A$, and $D\phi$ is the covariant derivative of an adjoint Higgs field $\phi$.  The background space is the unit ball model of hyperbolic space with curvature $-1$, which has metric
\begin{equation}\label{ballmet}
ds^2(\mathbf{H}^3)=\frac{4(dX_1^2+dX_2^2+dX_3^2)}{(1-R^2)^2},
\end{equation}
where $R^2=X_1^2+X_2^2+X_3^2$ and $R<1$.  This metric enters \eqref{bog} through the Hodge star.  Monopoles must satisfy the boundary condition $|\phi|^2=-\frac{1}{2}\text{Tr}\phi^2\to p^2$ as $R\to 1$, and the monopole charge $N\in \mathbf{Z}$ is the degree of the map $\phi|_{R=1}:S^2\to S^2$.  

If $2p\in \mathbf{Z}$, then the hyperbolic monopole is equivalent to a Euclidean instanton with charge $I=2pN$.  To understand this equivalence, we first change to half-space coordinates:
\[
X_1+iX_2=\frac{2x_1+2ix_2}{1+2r+r^2+x_1^2+x_2^2},\quad X_3=\frac{-1+r^2+x_1^2+x_2^2}{1+2r+r^2+x_1^2+x_2^2}.
\]
The metric \eqref{ballmet} becomes
\[
ds^2=\frac{dx_1^2+dx_2^2+dr^2}{r^2}.
\]
The metric on Euclidean $\mathbf{R}^4$ is
\[
ds^2=dx_1^2+dx_2^2+dx_3^2+dx_4^2.
\]
If we let $x_3+ix_4=re^{i\theta}$, then this becomes:
\[
dx_1^2+dx_2^2+dr^2+r^2d\theta^2=r^2\left(\frac{dx_1^2+dx_2^2+dr^2}{r^2}+d\theta^2\right),
\]
which shows that Euclidean $\mathbf{R}^4\setminus \mathbf{R}^2$ is conformal to $\mathbf{H}^3\times S^1$.  The Yang-Mills equations are conformally invariant, so if we have an instanton symmetric under rotations in the $\theta$-direction we can (after choosing a $\theta$-independent gauge) dimensionally reduce along the $\theta$-direction to give a solution to \eqref{bog}.  All our new monopole solutions are for the simplest case $p=1/2$, for which the monopole and instanton charge are equal. 
\section{Circle-invariant ADHM data}
The ADHM construction \cite{ADHM}, which we review in the next section, is an equivalence between instantons and quaternionic matrices satisfying algebraic constraints.  This means we can find hyperbolic monopoles by looking for ADHM data symmetric under a circle action.  Braam and Austin \cite{BA} analysed the circle-equivariant ADHM construction, and showed that one can write such data as a set of matrix difference equations defined on a lattice.  However, the Braam-Austin equations are still difficult to solve, presumably because they are based on a circle action that leads to the half-space model of hyperbolic space, so one cannot impose Platonic symmetries as in the Euclidean case \cite{HMM}.  In \cite{PHM}, Manton and Sutcliffe studied ADHM data invariant under a different circle action which leads to the ball model, and were able to obtain many solutions with commuting Platonic symmetries.  Below we review both the Manton-Sutcliffe and Braam-Austin approaches, as well as the spectral curves and rational maps one can associate to hyperbolic monopoles.
\subsection{Manton-Sutcliffe constraints}
The standard way of writing ADHM data is in terms of quaternionic matrices.  This uses an identification of $\mathbf{R}^4$ with the quaternions $\mathbf{H}$, so that a point $x\in \mathbf{R}^4$ is written as $x_1+ix_2+jx_3+kx_4$.  The ADHM data for a charge $I$ instanton is given by a pair of quaternionic matrices $L$ and $M$, where $L$ is an $I$-component row vector, and $M$ is a symmetric $I\times I$ matrix.  These are combined into
\[
\widehat{M}=\begin{pmatrix}
L\\
M
\end{pmatrix}.
\]
The ADHM constraint is
\begin{equation}\label{ADHMconstr}
\widehat{M}^\dagger\widehat{M}=R_I,
\end{equation}
where $R_I$ is a non-singular, real $I\times I$ matrix.  One then constructs the ADHM operator
\[
\Delta(x)=\begin{pmatrix}
L\\
M
\end{pmatrix}-\begin{pmatrix}
0\\
1_I
\end{pmatrix}x.
\]
To find the gauge field, one needs to find an $(I+1)$-component column vector $\Psi$ with $\Psi^\dagger\Psi=1$, that solves
\[
\Psi^\dagger\Delta(x)=0.
\]
The gauge potential is then given by
\[
A_\mu=\Psi^\dagger\partial_\mu\Psi,
\]
where the pure quaternion is regarded as an element of $su(2)$.

Now we consider circle-invariant ADHM data.  We can write conformal transformations of $\mathbf{R}^4$ as quaternionic M\"obius transformations
\[
x\to x^\prime=(Ax+B)(Cx+D)^{-1}
\]
The circle action Manton and Sutcliffe use is
\begin{equation}\label{MSca}
\begin{pmatrix}
A & B\\
C & D
\end{pmatrix}=
\begin{pmatrix}
\cos\frac{\alpha}{2} & \sin\frac{\alpha}{2}\\
-\sin\frac{\alpha}{2} & \cos\frac{\alpha}{2}
\end{pmatrix}
\end{equation}
The quotient of $\mathbf{R}^4$ by this action is the unit ball with metric conformal to \eqref{ballmet}.  The Manton-Sutcliffe constraints for an instanton to be invariant under this action are:
\begin{enumerate}\label{MScons}
\item $M$ is pure quaternion and symmetric,
\item $\widehat{M}^\dagger\widehat{M}=1_I$,
\item $LM=\mu L$, where $\mu$ is a pure quaternion, and $L$ is non-vanishing.
\end{enumerate}
ADHM data satisfying these constraints will correspond to a hyperbolic monopole with $p=1/2$.  All of the new monopole solutions in this paper will be given in terms of ADHM matrices satisfying the Manton-Sutcliffe constraints.

To calculate the Higgs field and energy density, suppose that the pure quaternion $X=X_1i+X_2j+X_3k$ represents a point in the unit ball.  Let $V(X)$ be a unit vector satisfying
\[
V^\dagger \Delta(X)=0,
\]
then the Higgs field is
\[
\phi=\frac{1}{2} V^\dagger \begin{pmatrix}
-\mu & L\\
-L^\dagger & M
\end{pmatrix}V,
\]
and the energy density is given by
\begin{equation}\label{ed}
\mathcal{E}=\frac{1}{\sqrt{g}}\partial_i (\sqrt{g}g^{ij}\partial_j|\phi|^2)
\end{equation}
where $g$ is the ball metric \eqref{ballmet}.

We will be interested in subgroups of the $SO(3)$ group of transformations of the form
\[
\begin{pmatrix}
A & B\\
C & D
\end{pmatrix}=\begin{pmatrix}
k & 0\\
0 & k
\end{pmatrix}
\]
where $k$ is a unit quaternion.  These transformations commute with the circle action \eqref{MSca} and will correspond, after dimensional reduction, to the group of rotations of $\mathbf{H}^3$ fixing the origin.  For an instanton to be symmetric under these transformations, we require
\begin{equation}\label{ADHMsyms}
\begin{pmatrix}
q & 0\\
0 & \mathcal{O}
\end{pmatrix}\begin{pmatrix}
L\\
M
\end{pmatrix}\mathcal{O}^{-1}=k\begin{pmatrix}
L\\
M
\end{pmatrix}k^{-1},
\end{equation}
where $q$ is a unit quaternion, and $\mathcal{O}\in O(I)$.  As $k$ runs over the elements of some symmetry subgroup of $SO(3)$, the corresponding matrices $\mathcal{O}(k)$ will furnish a real $I$-dimensional representation of the symmetry group, while $q(k)$ will give a 2-dimensional complex representation.
\subsection{The Braam-Austin construction}\label{BAconstr}
The circle action used in the Braam-Austin construction is
\[
\begin{pmatrix}
A & B\\
C & D
\end{pmatrix}=
\begin{pmatrix}
e^{i\theta/2} & 0\\
0 & e^{-i\theta/2}
\end{pmatrix}.
\]
The construction works for any $p$, and Braam and Austin showed that circle-equivariant ADHM data breaks up into a set of difference equations defined on a lattice with $2p$ sites.  For simplicity, we restrict to $2p$ odd.  For a monopole of charge $N$, the data consists of complex $N\times N$ matrices $\beta_i,\gamma_i$, and an $N$-row vector $v$.  The $\beta_i$ are defined on the even lattice points $i\in \{-2p+1,-2p+3,\dots,2p-1\}$, and the $\gamma_i$ are defined on the odd lattice points $i\in \{-2p+2,-2p+4,\dots,2p-2\} $.  This data must satisfy the Braam-Austin equations
\begin{align}\label{BA1}
\gamma_i-\gamma_{-i}^t&=0\\ \label{BA2}
\beta_i-\beta_{-i}^t&=0\\ \label{BA3}
\beta_{i-1}\gamma_i-\gamma_i\beta_{i+1}&=0\\\label{BA4}
[\beta_i^\dagger,\beta_i]+\gamma^\dagger_{i-1}\gamma_{i-1}-\gamma_{i+1}\gamma_{i+1}^\dagger&=0\\\label{BA5}
[\beta_{2p-1},\beta_{2p-1}^\dagger]+v^t \overline{v}-\gamma^\dagger_{2p-2}\gamma_{2p-2}&=0
\end{align}
Solutions to these difference equations correspond to a hyperbolic monopole in the half-space model.  

The Braam-Austin equations have a gauge freedom. Suppose that $g_i$ is a sequence of unitary matrices for $i\in\{-2p+1,-2p+3,\dots,2p-1\}$ with $g_i=\overline{g}_{-i}$.  Then it is easy to see that the gauge transformations
\begin{align}
\beta_i &\to g_i\beta_i g_i^{-1}\\
\gamma_i &\to g_{i-1}\gamma_i g_{i+1}^{-1}\\
v &\to vg^{-1}_{-2p+1}
\end{align}
leave the Braam-Austin equations invariant.  We also have the freedom to multiply $v$ by a unit norm complex number.  

The simplest case is $N=1$.  Here we can use the gauge freedom to set $\beta_i=\beta$ and $\gamma_i=\gamma=v$ independent of $i$, where $\beta$ is a complex number, and $\gamma$ is a positive real number.  $(\beta,\gamma)$ can then be interpreted as coordinates for the 1-monopole in the upper half-plane.  The $p=1/2$ case is also particularly simple.  In this case the Braam-Austin data just consists of a single complex matrix $\beta$ and a complex row vector $v$. If we identify these with standard form ADHM data: $L=v$, $M=\beta$, then the Braam-Austin conditions become the usual ADHM constraints \eqref{ADHMconstr}.

\subsection{Holomorphic data}
Just as for Euclidean monopoles, hyperbolic monopoles are known, via twistor correspondences, to be equivalent to spectral curves and rational maps between Riemann spheres.  

The spectral curve is a Riemann surface in the twistor space of oriented geodesics.  A geodesic lies on the spectral curve if the scattering equation
\begin{equation}\label{scat}
(D_s-i\phi)w=0
\end{equation}
along the geodesic has a normalisable solution.  In hyperbolic space, oriented geodesics are parametrised by their end-points on the boundary, and we denote the geodesic running from $\hat{\eta}=-1/\overline{\eta}$ (the antipodal point to $\eta$) to $\zeta$ by the pair $(\eta ,\zeta )\in \mathbf{CP}^1\times\mathbf{CP}^1$.  In these coordinates, the spectral curve of a charge $N$ monopole can be written as
\begin{equation}
\sum^N_{i=0,j=0}c_{ij}\eta^i\zeta^j=0,
\end{equation}
where the $c_{ij}$ are complex constants.

For $p=1/2$ Braam-Austin data, which is simply a complex ADHM matrix, the spectral curve is given by the formula
\begin{equation}
\det (\eta\zeta M^\dagger+\zeta-\eta\widehat{M}^\dagger\widehat{M}-M)=0.
\end{equation}
Atiyah discovered a correspondence between hyperbolic $N$-monopoles and based rational maps of degree $N$, which is also defined using the scattering equation \eqref{scat}.  One considers geodesics running from $\hat{\eta}=\infty$ to $\zeta=z$, and $\mathcal{R}(z)$ is defined to be the ratio of the decaying to the growing component of the solution to \eqref{scat} at the $\zeta=z$ end.  The basing condition is that $\mathcal{R}(\infty)=0$, so the numerator of $\mathcal{R}$ has degree less than $N$, and the denominator has degree $N$.  This correspondence is not quite one-to-one, because multiplying $\mathcal{R}(z)$ by a constant phase corresponds to an identical monopole.  We shall call these Donaldson-type maps, since they are the hyperbolic analogue of the correspondence between Euclidean monopoles and rational maps discovered by Donaldson \cite{D}.  For $p=1/2$ Braam-Austin data, we have a simple formula for the Donaldson-type rational map \cite{MS}:
\begin{equation}\label{Don}
\mathcal{R}(z)=L(z-M)^{-1}L^t.
\end{equation}

Jarvis defined a rational map for Euclidean monopoles more adapted to rotational symmetries than the Donaldson map \cite{J}.  In the hyperbolic case, if we have ADHM data satisfying the Manton-Sutcliffe constraints for circle-invariance, as well as $\mu=0$,  a formula for a rational map which appears to be of Jarvis type was given in \cite{PHM}:
\begin{equation}\label{Jar}
f(X)=L(M-X)^{-1}L^\dagger,
\end{equation}
where $X$ is a unit pure quaternion representing a point on the boundary.  One obtains a rational map by writing both $X$ and its image $f(X)$ in Riemann sphere coordinates.  For all the known examples of monopoles in the Manton-Sutcliffe formalism, the rational map \eqref{Jar} has the same symmetry as the corresponding monopole.  We shall see that this is also true of all the monopoles in this paper with $\mu=0$.

\section{Axial hyperbolic monopoles}
\subsection{Axial hyperbolic monopoles from the JNR ansatz}
In this section we shall derive explicit expressions for the the fields and ADHM data of axially symmetric $p=1/2$ monopoles.  Before discussing this, we first review the construction of axial monopoles from the JNR ansatz.

The JNR ansatz \cite{JNR} gives a charge $I$ instanton by specifying $I+1$ points $\{a_j\}$ in $\mathbf{R}^4$, together with $I+1$ positive real numbers $\{ \lambda_j\}$.  The gauge field is then
\begin{equation}
A_\mu=\frac{i}{2}\sigma_{\mu\nu}\,\partial_\nu\log\left( \sum^I_{j=0}\frac{\lambda^2_j}{|x-a_j|^2} \right),
\end{equation}
where $\sigma_{i4}=\tau_i$, $\sigma_{ij}=\epsilon_{ijk}\tau_k$, and $\tau_i$ $(i=1,2,3)$ are the Pauli matrices.

If one places the poles $a_j$ on the fixed plane of some circle action on $\mathbf{R}^4$, then the corresponding instanton will necessarily be invariant under the circle action, and so correspond to a hyperbolic monopole.  To obtain a monopole with the same symmetry as the configuration of poles, one must set the poles to have equal weight after a conformal transformation to the unit ball.  If the poles lie on a circle then there is an action of the conformal group rotating the poles and acting on their weights, which means that the monopole can have extra symmetry not present in the configuration of poles.  In particular, placing the poles at the vertices of a regular $n$-gon gives a symmetry enhancement from $\mathbf{Z}_n$ to $U(1)$.  This gives a straightforward way to construct axially symmetric hyperbolic monopoles.

The JNR construction readily gives the fields for hyperbolic monopoles and can be straightforwardly written in standard ADHM matrix form using the methods described in \cite{BCS}.  Unfortunately the resulting ADHM data will not be in a tractable form, and this approach does not give the representations (as in \eqref{ADHMsyms}) compensating for the axial symmetry.  These representations will be important later when we try to deform the axial monopoles to give families that are not obtainable from the JNR ansatz.
\subsection{Axial monopoles and 1-monopoles}\label{swap}
Our first result uses a correspondence between axial $p=1/2$ $N$-monopoles and $p=N/2$ 1-monopoles to derive explicit axial monopole fields.  To see this correspondence, start with a $p=N/2$ 1-monopole.  This monopole is equivalent to a charge $N$ instanton invariant under rotations in the $x_3x_4$-plane. However, 1-monopoles have an $SO(3)$-symmetry group of rotations about their centres, and in particular are symmetric under the $SO(2)$ subgroup of rotations in the $x_1x_2$ plane.  This $SO(2)$ symmetry lifts to the underlying instanton.  The idea is now to swap the roles of these two symmetries, so we quotient by rotations in the $x_1x_2$-plane and view rotations in the $x_3x_4$-plane as a physical symmetry of the resulting monopole.  The axially symmetric monopole one obtains after this swap will have charge $N$ and $p=1/2$.

This observation is useful because 1-monopoles are particularly simple.  The Bogomolny equation can be solved \cite{N} by a spherically-symmetric ansatz for all values of $p$:
\begin{equation}\label{spher}
A_i^a=\frac{2(P(R)-1)}{R^2}\epsilon_{iak}X^k\textrm{ and } \phi^a=\frac{Q(R)X_a}{R}
\end{equation}
where
\[
P(R)=\frac{B\sinh s}{\sinh Bs }\textrm{, }Q(R)=\coth s-B \coth Bs
\]
and $s=2\tanh^{-1}R$, $B=2p+1$.  If $p$ is a half-integer, then $P$ and $Q$ are rational functions of $R$.

To obtain axial monopole fields, one first performs a coordinate transformation of the fields \eqref{spher} to half-space $(A_{X_1},A_{X_2},A_{X_3},\phi)\to (A_1,A_2,A_r,\phi)$.  To lift to the instanton, we interpret $(A_r,\phi)$ as radial and angular components respectively of $A$ in the $x_3x_4$-plane.  This instanton is symmetric under rotations in the $x_1x_2$ plane, but to dimensionally reduce along this direction we must first put it in a gauge in which it is independent of these rotations.  This process gives the Higgs field magnitude of an axial $2p$-monopole, written in ball model coordinates:
\begin{equation}
|\phi|^2=(P(S)-1)^2 \frac{(1-R^2)^2(R^2-\rho^2)}{((1+R^2)^2-4\rho^2)^2}+\frac{1}{4}\left((P(S)-1)\frac{(1-R^2)^2}{(1+R^2)^2-4\rho^2}+1\right)^2,
\end{equation}
where $\rho^2=X_1^2+X_2^2$ and $S=2\sqrt{\frac{1+R^2-2\rho}{1+R^2+2\rho}}$.
\subsection{Axial monopole ADHM data from the Braam-Austin construction}\label{axba }
The real usefulness of this `swap' of the roles of the circle symmetries is that it gives the axial monopole ADHM data in a simple, natural form. Carrying out the swap at the ADHM level means that the Braam-Austin data for a 1-monopole written as a standard-form ADHM matrix is precisely the same as the Braam-Austin data for the corresponding axial monopole.  Braam-Austin data for a 1-monopole is trivial to write down, so the matrix one obtains this way is much simpler than the one coming from the JNR ansatz.  Furthermore one easily obtains the matrices compensating for the axial symmetry.  Below we give the results of this construction;  full details of the derivation can be found in the Appendix.

The ADHM data for the axial charge $N$, $p=1/2$ monopole is best given inductively.  The data for the 2-monopole is
\begin{equation}\label{c2}
\widehat{M}^{\text{ax}}_{2}=\frac{1}{2}\begin{pmatrix}
j\sqrt{2} & -k\sqrt{2}\\
-j & -k\\
-k & j
\end{pmatrix}
\end{equation}
while the data for the 3-monopole is
\begin{equation}\label{c3} \widehat{M}^{\text{ax}}_{3}=\frac{1}{\sqrt{2}}\begin{pmatrix}
j & 0 & -k\\
0 & j & 0\\
j & 0 & k\\
0 & k & 0
\end{pmatrix}.
\end{equation}
Now suppose that
\[
\widehat{M}^{\text{ax}}_{N}=\begin{pmatrix}
L^\text{ax}_{N}\\
M^\text{ax}_{N}
\end{pmatrix}
\]
is the ADHM data for an axial $N$-monopole for $N\geq 2$.  Then
\begin{equation}\label{gen}
\widehat{M}_{N+2}^\text{ax}=\begin{pmatrix}
j/\sqrt{2} & 0 & 0 & \cdots & 0 & 0 & -k/\sqrt{2}\\
0 & j/2 & 0 & \cdots & 0 & -k/2 & 0\\
j/2 &    &   & &     &    & k/2\\
0 & &		&	& & & 0\\
\vdots & & 	& M_{N}^\text{ax}	& & & \vdots\\
0 & 	&		&	& & & 0\\
-k/2 & 			 &		& &	& & j/2\\
0 & k/2 & 0 & \cdots & 0 & j/2 & 0
\end{pmatrix}
\end{equation}
For axial symmetry, the matrices $\widehat{M}_{N}^\text{ax}$ satisfy
\begin{equation}\label{axsym}
\begin{pmatrix}
q_{N} & 0\\
0 & \mathcal{O}_{N}
\end{pmatrix}\widehat{M}_{N}^\text{ax}\mathcal{O}^{-1}_{N}=e^{-i\theta}\widehat{M}^\text{ax}_{N}
\end{equation}
where 
\begin{equation}\label{qrep}
q_{N}(\theta)=e^{-i\theta(N+1)/2}
\end{equation}
and we define $\mathcal{O}_{N}$ inductively. Firstly,
\[
\mathcal{O}_1(\theta)=1\textrm{ and }\mathcal{O}_{2}(\theta)=\begin{pmatrix}
\cos\theta/2 & -\sin\theta/2\\
\sin\theta/2 & \cos\theta/2
\end{pmatrix}
\]
and, for $N\geq 3$,
\begin{equation}\label{rep}
\mathcal{O}_{N+2}(\theta)=
\begin{pmatrix}
\cos{(N+1)\theta/2} & 0 &\cdots & 0 & -\sin{(N+1)\theta/2}\\
0 & & & & 0\\
\vdots & &\mathcal{O}_{N}(\theta) & & \vdots\\
0 & & & & 0\\
  \sin{(N+1)\theta/2} & 0 &\cdots  & 0 & \cos{(N+1)\theta/2}
\end{pmatrix}
\end{equation}
One can check that the matrices defined by \eqref{gen} satisfy the Manton-Sutcliffe constraints for circle symmetry.  This is actually unsurprising.  We would obtain the same instantons by placing the poles of the JNR ansatz at the roots of unity with equal weight in the $jk$-plane.  But then the poles will lie on the fixed-point set of the Manton-Sutcliffe circle action, which is the 2-sphere of unit-norm pure imaginary quaternions, so the instanton must also be invariant under this action as well.

In the next section we will be interested in subgroups of the symmetry group of the axial monopoles.  Axial monopoles are invariant under reflection in the $X_2X_3$-plane
\[
I:X\to iXi
\]
since $i\widehat{M}_N i=\widehat{M}_N$, and the compensating transformation is just the identity.  Axial $N$-monopoles are also symmetric under rotations by $\pi$ around the $X_2$-axis
\[
R_2:X\to -jXj
\]
since
\begin{equation}
-j\widehat{M}_Nj=\begin{pmatrix}
1 & 0\\
0 & \mathcal{O}^R_N
\end{pmatrix}\widehat{M}_N(\mathcal{O}^R_N)^{-1}
\end{equation}
where
\[
(\mathcal{O}^R_N)_{ab}=\begin{cases} 1 & \text{if $a=b$ and $1\leq a\leq (N+1)/2$}\\
-1 & \text{if $a=b$ and $(N+1)/2 < a\leq N$}\\
0 & \text{otherwise}
\end{cases}
\]
For reference, the axial $N$-monopole spectral curve is
\begin{equation}
\mathcal{A}_N(\eta,\zeta)\equiv\sum^N_{i=0} (-1)^i\eta^i\zeta^{N-i}=0
\end{equation}
and the Jarvis and Donaldson-type rational maps are both $1/z^N$ .
\section{Deforming the axial monopole}
In this section we derive circle-invariant ADHM data corresponding to families of dihedral and twisted-line symmetric hyperbolic monopoles.  Our data will satisfy the Manton-Sutcliffe constraints for circle invariance.  To find symmetric ADHM data, one can take representations of the symmetry qroups in $\mathbf{C}^2$ and $O(N)$, and use them as constraints on ADHM data via \eqref{ADHMsyms}.  For low charges, solving for ADHM data constrained by the dihedral and twisted line symmmetries as well as the Manton-Sutcliffe conditions is tractable, and the solutions can be easily generalised to higher charges.  This section contains the results of this approach for some interesting symmetry groups.  The representations \eqref{rep} are key for this derivation, since we are using subrepresentations of these to constrain the data.
\subsection{$D_N$-symmetric $N$-monopoles}\label{dnsymn}
The dihedral group $D_N$ is generated by rotations by $2\pi/N$ around the $X_1$-axis, and the rotation $R_2$ given above.  Using the representations of these symmetries given in section \ref{axba } leads to the following family of $D_N$-symmetric $N$-monopoles, for $N>2$:
\begin{equation}\label{cyc}
\widehat{M}_{N}=\frac{1}{2}\begin{pmatrix}
j \sqrt{2(1-\alpha^2)} & 0 & & \cdots & & 0 & -k\sqrt{2(1-\alpha^2)}\\
(-1)^Nj\alpha & j & 0 & \cdots & 0 & -k & (-1)^{N+1}k\alpha\\
j & & & & & & k\\
0 & & & & & & 0\\
\vdots & & & 2M^\text{ax}_{N-2} & & & \vdots\\
0 & & & & & & 0\\
-k & & & & & & j\\
(-1)^{N+1}k\alpha & k & 0 & \cdots & 0 & j & (-1)^{N+1}j\alpha
\end{pmatrix}
\end{equation}
for $\alpha\in(-1,1)$.  It is straightforward to check that the ADHM data \eqref{cyc} also satisfies the Manton-Sutcliffe constraints, and in particular that $\mu=(-1)^N\alpha j$ for all $N$.

These families correspond to the simplest type of $N$-monopole scattering.  The sign of $\alpha$ is chosen so that $N$ monopoles are close to the roots of unity for $\alpha$ close to 1.  As $\alpha$ decreases, the monopoles move radially towards the origin, losing their individual identities as $\alpha$ gets closer to 0.  The configuration becomes axial at $\alpha=0$, and as $\alpha$ becomes negative individual monopoles re-emerge in the same configuration as the incoming ones, but rotated by $\pi/N$.  The corresponding geodesic for Euclidean monopoles is known only from its Donaldson rational maps.

If we multiply by $j$ on the left, then the data \eqref{cyc} is purely complex, so we can think of it as Braam-Austin data defining a monopole in the half-plane model.  We can calculate spectral curves and rational maps in this setting.  For $N=3$, the spectral curve is
\begin{equation}\label{d3sc}
\eta^3-\zeta^3-\alpha(\eta^3\zeta^3-1)+(\alpha^2-1)(\eta^2\zeta-\eta\zeta^2)=0,
\end{equation}
which has manifest symmetry under $(\eta,\zeta)\to (e^{2i\pi/3}\eta,e^{2i\pi/3}\zeta)$.  This curve is also symmetric under the map $(\eta,\zeta)\to(1/\eta,1/\zeta)$, so the monopole has dihedral symmetry in both the ball and half-space models.  Indeed, if we make the identification $\alpha=-a$, then \eqref{d3sc} is the same as the spectral curve of the $D_3$-symmetric 3-monopole family discussed in \cite{BCS}.  This explains why the data \eqref{cyc} for $N=3$ gives $D_3$-symmetric monopoles in both models: just as for the axial monopole, the JNR poles lie on the fixed-point sets of both circle actions.  These $D_N$-symmetric $N$-monopoles appear to lie within the space of JNR data for all $N$, although the required configuration of poles and weights is rather complicated.  Here these monopoles appear as simple, natural deformations of the axial monopole ADHM data.  The Donaldson-type rational maps corresponding to \eqref{cyc} are
\begin{equation}\label{dmap}
\mathcal{R}(z)=\frac{1-\alpha^2}{z^N+\alpha}.
\end{equation}
We can prove these maps straightforwardly using a formal expansion in powers of $z^{-1}$.  The expansion of \eqref{dmap} is
\[
\sum^\infty_{j=0} (1-\alpha^2)\frac{(-\alpha)^j}{z^{N(j+1)}
}
\]
while the expansion of the general rational map formula is
\[
L(z-M)^{-1}L^t=\sum^\infty_{k=0}\frac{LM^kL^t}{z^{k+1}},
\]
where for the rest of this section $L$ and $M$ are the top row and bottom $N$ rows respectively of \eqref{cyc} multiplied on the left by $j$.  The coefficients $LM^kL^t$ can be calculated explicitly.  Note that for $1\leq k<N/2-1$,
\[
(0, \dots, 0 ,\underbrace{1}_{k\textrm{-th}}, 0 ,\dots,0 ,\underbrace{-i}_{(N-k+1)\text{-th}} , \dots , 0)
M_N=(0, \dots, 0 ,\underbrace{1}_{(k+1)\text{-th}}, 0 ,\dots,0 ,\underbrace{-i}_{(N-k)\text{-th}} , \dots , 0)
\]
so we have, for $0\leq k\leq N/2-1$,
\[
LM^k=\sqrt{\frac{1-\alpha^2}{2}}(0, \dots, 0 ,\underbrace{1}_{(k+1)\textrm{-th}}, 0 ,\dots,0 ,\underbrace{-i}_{(N-k)\text{-th}} , \dots , 0).
\]
Similarly, for $N/2\leq k\leq N-1$, if $N$ is even,
\[
LM^{k}=\sqrt{\frac{1-\alpha^2}{2}}(0, \dots, 0 ,\underbrace{-1}_{(N-k)\textrm{-th}}, 0 ,\dots,0 ,\underbrace{-i}_{(k+1)\text{-th}} , \dots , 0),
\]
and if $N$ is odd, 
\begin{align*}
LM^{(N-1)/2}=\sqrt{\frac{1-\alpha^2}{2}}(0, \dots, 0 ,\underbrace{1}_{(N+1)/2\textrm{-th}}, 0 ,\dots,0)&\\
LM^{k}=\sqrt{\frac{1-\alpha^2}{2}}(0, \dots, 0 ,\underbrace{1}_{(N-k)\textrm{-th}}, 0 ,\dots,0 ,\underbrace{i}_{(k+1)\text{-th}} , \dots , 0)&\text{\quad for $(N+1)/2\leq k\leq N-1$}.
\end{align*}
This shows that $LM^kL^t=0$ for $0\leq k<N-1$, and $LM^{N-1}L^t=(-1)^{N}(\alpha^2-1)$.  Also
\[
LM^N=-\alpha\sqrt{\frac{1-\alpha^2}{2}}(1,0,\dots,0,-i)=-\alpha L,
\]
and so 
\[
LM^kL^t=\begin{cases}
0 \quad \text{if $(k+1)$ mod $N\neq 0$}\\
(-1)^N(-\alpha)^{(k+1)/N-1}(\alpha^2-1)\quad\text{if $(k+1)$ mod $N= 0$.}
\end{cases}
\]
so the coefficients of the two expansions are equal up to an overall phase of $(-1)^N$, proving the rational map formula \eqref{dmap}.  The rational map formula also makes the cyclic symmetry $z\to e^{2\pi i/N}z$ manifest.

The generalisation of the spectral curve formula \eqref{d3sc} to arbitrary $N$ appears to be
\begin{equation}\label{dnsc}
\mathcal{A}_N+\alpha (1-(-\eta\zeta)^N)+\alpha^2(\zeta^N+(-\eta)^N-\mathcal{A}_N)=0,
\end{equation}
although we have not been able to prove this formula from the data \eqref{cyc}.
\subsection{$D_{N-j}$-symmetric $N$-monopoles}
We can generate another interesting family by imposing $D_{N-1}$ symmetry together with the constraint $\mu=0$.  For $N>2$, the resulting ADHM data is
\begin{equation}\label{d1sym}
\frac{1}{2}\begin{pmatrix}
j \sqrt{2-2(-1)^N\alpha} & 0 & 0 & \cdots & 0 & 0 & -k\sqrt{2+2(-1)^N\alpha}\\
0 & j\sqrt{1+(-1)^N\alpha} & 0 & \cdots & 0 & -k\sqrt{1+(-1)^N\alpha} & 0\\
j\sqrt{1+(-1)^N\alpha} &  &  &  &  & & k\sqrt{1-(-1)^N\alpha}\\
0 & & & & & & 0\\
\vdots & &  & 2M^\text{ax}_{N-2} & & & \vdots\\
0 & & & & & & 0\\
-k\sqrt{1+(-1)^N\alpha} &  &  &  &  & & j\sqrt{1-(-1)^N\alpha}\\
0 & k\sqrt{1-(-1)^N\alpha} & 0 & \cdots & 0 & j\sqrt{1-(-1)^N\alpha} & 0
\end{pmatrix}
\end{equation}
for $\alpha\in (-1,1)$.  For $\alpha$ close to 1, the configuration consists of $N-1$ monopoles at the roots of unity, and one monopole at the origin.  As $\alpha$ decreases, the outer monopoles approach the origin from infinity, while a 1-monopole stays at the origin throughout. The configuration becomes axial for $\alpha=0$, and as $\alpha$ becomes negative the same process happens in reverse, with the configuration rotated by an angle of $\pi/(N-1)$.  The first column of Figure 1 shows  energy density isosurfaces of $D_2$ symmetric 3-monopoles at several different values of $\alpha$ between 1 and $-1$, decreasing down the column.

This family can be generalised to give families with two or three monopoles at the origin.  Other families with more than three at the origin must exist, but we have not been able to find them with our methods.  All these $D_{N-j}$-symmetric $N$-monopole families follow the same pattern as the $D_{N-1}$-symmetric $N$-monopole families.  For $\alpha$ close to 1, $(N-j)$ 1-monopoles are arranged in a polygon around an axial $i$-monopole at the origin.  As $\alpha$ decreases, the 1-monopoles move radially inwards, `scattering' through the axial configuration at $\alpha=0$, and coming out again rotated by an angle of $\pi/(N-j)$.

The data corresponding to a $D_2$-symmetric 4-monopole is
\begin{equation}\label{d2c4}
\frac{1}{2}
\begin{pmatrix}
j \sqrt{2} & 0 & 0 & -k \sqrt{2}\\
-j\alpha & j\sqrt{1-\alpha^2} & -k\sqrt{1-\alpha^2} & -k\alpha\\
j\sqrt{1-\alpha^2} & j(-1+\alpha) & k(-1-\alpha) & k\sqrt{1-\alpha^2}\\
-k\sqrt{1-\alpha^2} & k(-1-\alpha) & j(1-\alpha) & j\sqrt{1-\alpha^2}\\
-k\alpha & k\sqrt{1-\alpha^2} & j\sqrt{1-\alpha^2} & j\alpha
\end{pmatrix}
\end{equation}
The second column of Figure 1 shows energy density isosurfaces for this family for different values of $\alpha$ between 1 and $-1$, decreasing down the column.

For $N>4$, the data corresponding to a $D_{N-2}$-symmetric $N$-monopole is 
\begin{equation}\label{dn2cn}
\frac{1}{2}\begin{pmatrix}
j\sqrt{2} & 0 &  & & \dots & &  & 0 & -k\sqrt{2}\\
(-1)^{N+1}j\alpha & j\sqrt{1-\alpha^2} & 0 & & \dots & & 0 & -k\sqrt{1-\alpha^2} & (-1)^{N+1}k\alpha\\
j\sqrt{1-\alpha^2} & (-1)^{N}j\alpha & j & 0 & \dots & 0 & -k & (-1)^{N+1}k\alpha & k\sqrt{1-\alpha^2}\\
0 & j & & & &  &  & k & 0\\
0 & 0 & & & & & & 0 & 0\\
\vdots & \vdots & & & 2M_{N-4}^{\text{ax}} & & & \vdots & \vdots\\
0 & 0 & & & & & & 0 & 0\\
0 & -k & & & & &  & j & 0\\
-k\sqrt{1-\alpha^2} & (-1)^{N+1}k\alpha & k & 0 & \dots & 0 & j & (-1)^{N+1}j\alpha & j\sqrt{1-\alpha^2}\\
(-1)^{N+1}k\alpha & k\sqrt{1-\alpha^2} & 0 & & \dots & & 0 & j\sqrt{1-\alpha^2} & (-1)^{N}j\alpha\\
\end{pmatrix}
\end{equation}
For $N>4$, the data corresponding to a $D_{N-3}$-symmetric $N$-monopole is
\begin{equation}\label{dn3n}
\frac{1}{2}\begin{pmatrix}
j\sqrt{2} & 0 &  & & \dots & &  & 0 & -k\sqrt{2}\\
0 & j\sqrt{1-\alpha} & 0 & & \dots & & 0 & -k\sqrt{1+\alpha} & 0\\
j\sqrt{1-\alpha} & 0 & j\sqrt{1+\alpha} & 0 & \dots & 0 & -k\sqrt{1+\alpha} & 0 & k\sqrt{1-\alpha}\\
0 & j\sqrt{1+\alpha} & & & &  &  & k\sqrt{1-\alpha} & 0\\
0 & 0 & & & & & & 0 & 0\\
\vdots & \vdots & & & 2M_{N-4}^{\text{ax}} & & & \vdots & \vdots\\
0 & 0 & & & & & & 0 & 0\\
0 & -k\sqrt{1+\alpha} & & & & &  & j\sqrt{1-\alpha} & 0\\
-k\sqrt{1+\alpha} & 0 & k\sqrt{1-\alpha} & 0 & \dots & 0 & j\sqrt{1-\alpha} & 0 & j\sqrt{1+\alpha}\\
0 & k\sqrt{1-\alpha} & 0 & & \dots & & 0 & j\sqrt{1+\alpha} & 0\\
\end{pmatrix}
\end{equation}
For clarity we have left out the alternating sign of $\alpha$ in \eqref{dn3n}. Replacing $\alpha$ by $(-1)^N\alpha$ in \eqref{dn3n} would ensure that the outer monopoles lie near the roots of unity for $\alpha$ close to 1.

As in section \ref{dnsymn}, we can think of all the data in this section as defining monopoles in either the half-space or ball models.  The generalisation of the spectral curve \eqref{dnsc} to $D_{N-j}$-symmetric $N$-monopoles appears to be, for $0\leq j\leq 3$,
\begin{equation}\label{genax}
\mathcal{A}_N-\alpha \mathcal{A}_j\left((-1)^{j+1}+(-1)^N(\eta\zeta)^{N-j}\right)+
\alpha^2\left(\mathcal{A}_{j}(\zeta^{N-j}+(-\eta)^{N-j})-\mathcal{A}_N\right)=0.
\end{equation}
One can check that as $|\alpha|\to 1$ this becomes a product of $\mathcal{A}_j$ and stars for monopoles arranged in a regular polygon on the boundary.  This curve has the  rotation symmetries $(\eta,\zeta)\to(e^{2\pi i/(N-j)}\eta,e^{2\pi i/(N-j)}\zeta)$ and $(\eta,\zeta)\to (1/\eta,1/\zeta)$, so the data \eqref{d1sym}, \eqref{d2c4}, \eqref{dn2cn}, \eqref{dn3n}, all correspond to $D_{N-j}$-symmetric $N$-monopoles in either the half-space or ball models.  The corresponding Donaldson-type rational maps are
\[
\mathcal{R}(z)=\frac{1+\alpha z^{N-j}}{\alpha z^j+z^N}
\]
with manifest rotational symmetry $\mathcal{R}(e^{2\pi i/(N-j)}z)=e^{-2\pi i/(N-j)}\mathcal{R}(z)$.  These can be proved in the same way as the $D_N$-symmetric $N$-monopole rational map \eqref{dmap}.  If $j>0$, then $\mu=0$ for all values of $N$ and $\alpha$, so we can calculate the Jarvis rational maps for these families.  Interestingly, if one replaces $i\to k, j\to i, k\to j$ in the data above to ensure the monopole is in the correct orientation, then the Jarvis-type rational maps are precisely the same as the Donaldson-type maps.  The other $D_{N-j}$ symmetry generator manifests itself as $\mathcal{R}(1/z)=1/\mathcal{R}(z)$.
\subsection{Twisted-line symmetric monopoles}
We can also consider monopoles invariant under a `twisted inversion symmetry'.  The symmetry $I_N$ acts by the reflection $I$ in the $X_2X_3$ plane combined with a rotation of $2\pi/N$ around the $X_1$-axis:
\[
I_N:X\to ie^{\frac{2i\pi}{N}}Xe^{-\frac{2i\pi}{N}}i.
\]
We shall consider $N$-monopoles invariant under an $I_{2N-2}$ symmetry.  Such monopoles were originally considered in the Euclidean space context in \cite{HS}.
The first example is a 3-monopole with $I_4$ symmetry, given by the matrix
\begin{equation}\label{tl3}
\frac{1}{\sqrt{2}}\begin{pmatrix}
j\sqrt{1-\alpha^2} & i\sqrt{2} \alpha & -k\sqrt{1-\alpha^2}\\
i\sqrt{2}\alpha & j\sqrt{1-\alpha^2} & 0\\
j\sqrt{1-\alpha^2} & 0 & k\sqrt{1-\alpha^2}\\
0 & k\sqrt{1-\alpha^2} & -i\sqrt{2} \alpha
\end{pmatrix}
\end{equation}
for $\alpha\in (-1,1)$.  This family was described from the JNR point of view in \cite{BCS}.  The process consists of 3 monopoles scattering along the $X_1$-axis.  As $\alpha$ decreases from 1, two monopoles approach a central monopole at the origin.  When $\alpha=1/\sqrt{3}$, \eqref{tl3} becomes equal to (a rotated version of) the tetrahedrally symmetric ADHM data given in \cite{PHM}.  As $\alpha$ becomes negative, the same process happens in reverse, but the configuration is rotated by $\pi/2$ around the $X_1$ axis, so $\alpha=-1/\sqrt{3}$ corresponds to the dual tetrahedral configuration.

\begin{figure}
\begin{center}
\includegraphics[scale=0.6]{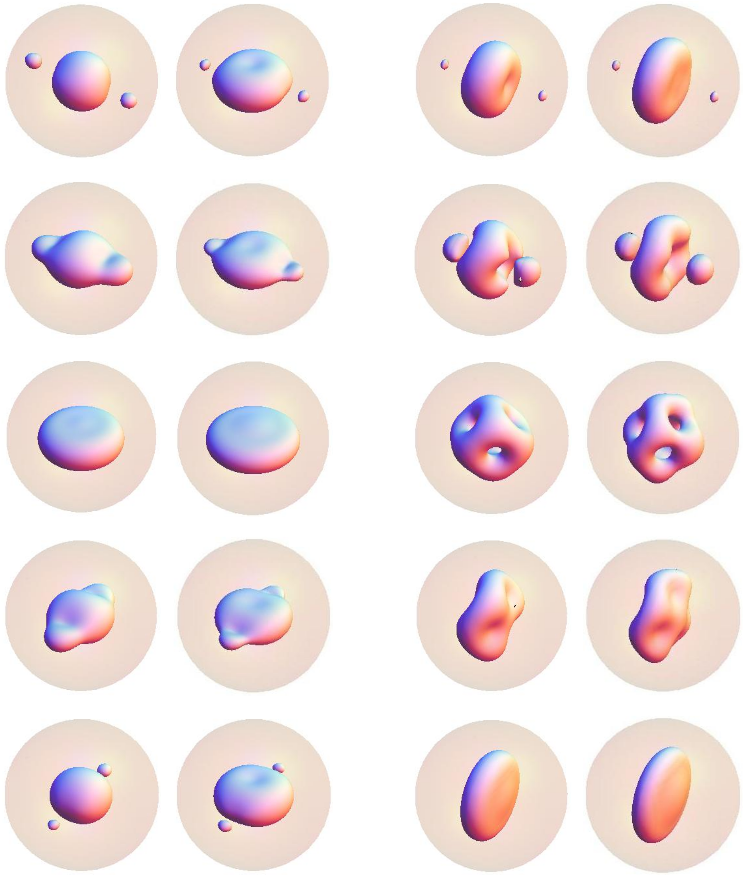}
\end{center}
Figure 1: Energy density isosurfaces: first column $D_2$-symmetric 3-monopoles, second column $D_2$-symmetric 4-monopoles, third column $I_6$-symmetric 4-monopoles, fourth column $I_8$-symmetric 5-monopoles.  The energy densities were calculated using the formula \eqref{ed}.
\end{figure}
The generalisation of \eqref{tl3} to all $N$ is
\begin{equation}\label{Iadhm}
\widehat{M}^I_{N}=\frac{1}{2}\begin{pmatrix}
j \sqrt{2-2\alpha^2} & -i\alpha\sqrt{2} & 0 & \cdots & 0 & \alpha\sqrt{2} & -k\sqrt{2-2\alpha^2}\\
-2i\alpha & j\sqrt{1-\alpha^2} & 0 & \cdots & 0 & -k\sqrt{1-\alpha^2} & 0\\
j\sqrt{1-\alpha^2} &  &  &  &  & & k\sqrt{1-\alpha^2}\\
0 & & & & & & 0\\
\vdots & &  & 2M_{N-2}^\text{ax} & & & \vdots\\
0 & & & & & & 0\\
-k\sqrt{1-\alpha^2} &  &  &  &  & & j\sqrt{1-\alpha^2}\\
0 & k\sqrt{1-\alpha^2} & 0 & \cdots & 0 & j\sqrt{1-\alpha^2} & 2i\alpha
\end{pmatrix}
\end{equation}
As $\alpha$ varies, the process consists of two 1-monopoles moving along the $X_1$-axis towards a central approximately axial cluster of $(N-2)$ monopoles at the origin.  In contrast to the dihedral families, the outer monopoles lie on the symmetry axis of the central cluster, rather than perpendicular to it. The configuration becomes axial as $\alpha$ passes through zero, and then as $\alpha$ decreases the process happens in reverse, but rotated by $\pi/(N-1)$ around the $X_1$-axis.  

Since $\mu=0$ for all members of the family \eqref{Iadhm}, we can calculate their Jarvis rational maps.  We again substitute $i\to k, j\to i, k\to j$ in the data to put the rational maps into a simpler form.  For $N=3$, the rational map is
\begin{equation}\label{i3rat}
\frac{2\alpha z^2+1-\alpha^2}{(1-\alpha^2)z^3-2\alpha z}
\end{equation}
which has the $I_4$ symmetry $R(i/\overline{z})=i/\overline{R(z)}$.  For $N=4$, the rational map is
\begin{equation}\label{i4rat}
\frac{2\alpha z^3+1-\alpha^2}{(1-\alpha^2)z^4-2\alpha z}
\end{equation}
with the $I_6$ symmetry $R(e^{i\pi/3}/\overline{z})=e^{i\pi/3}/\overline{R(z)}$.  The third and fourth columns of Figure 1 show energy density isosurfaces of twisted-line symmetric monopoles of charges 4 and 5 respectively, for different values of $\alpha$ between 1 and 0 decreasing down the column.

For the $N=4$ family, one can check that if $\alpha=\pm 1/\sqrt{2}$ the rational map \eqref{i4rat} is equivalent to the degree 4 cubically symmetric rational map given in \cite{HMS}, combined with a rotation to bring the cube into the correct orientation.  Note that for $N>3$, all members of these twisted-line symmetric families (apart from the axial monopoles) lie outside the space of JNR-type monopoles, because it is impossible to arrange $N+1$ distinct points on the 2-sphere with $I_{2N-2}$-symmetry.

\section{Conclusion}
Using an equivalence between $p=1/2$ axial $N$-monopoles and $p=N/2$ 1-monopoles, we have given an explicit formula for the Higgs field magnitude of axial $p=1/2$ monopoles, as well as their circle-invariant ADHM data.  We have deformed the axial monopole data to give 1-parameter families with various kinds of dihedral and twisted-line symmetry, for all values of the topological charge.

The dihedral families presented here should extend to $D_{N-j}$-symmetric $N$-monopole configurations with $j$ monopoles at the origin for $j>3$, although we have not been able to find the explicit ADHM data for these with our methods.  Another approach would be to see if there exist JNR-type monopoles with the right poles and weights to give the conjectured spectral curves \eqref{genax}.  For $j$ and $(N-j)$ both large, this should give a hyperbolic prototype of Manton's multi-shell magnetic bags \cite{M}, although these bags would be degenerate, in the sense that the volume of their interiors would be zero.

The natural $L^2$ metric on the moduli space of hyperbolic monopoles diverges, so the usual geodesic approximation to monopole scattering cannot be applied in this case.  However, hyperbolic monopoles define an abelian connection on the boundary 2-sphere, and this connection can be used to define a metric on the monopole moduli space.  An integral expression for this metric on the space of JNR-type monopoles was given in \cite{BCS}.  All the 1-parameter families in this paper are geodesics with respect to this metric, since they are obtained as fixed-point sets of subgroups of the symmetry group of hyperbolic space.  In this paper we gave the data for a family of $D_2$-symmetric 3-monopoles.  It would be interesting to compare the metric on this space with the metric on the space of $D_2$-symmetric 2-monopoles, since in the Euclidean case these are known to be the same.

\appendix
\section{Appendix}

In this appendix we derive the monopole ADHM data \eqref{gen} and the representations \eqref{rep} by interchanging the roles of the circle symmetries, as described in section \ref{swap}. To implement this `swap' at the ADHM level, one has to write the ADHM construction in the abstract, coordinate-free way it was originally introduced, and later used by Braam and Austin \cite{BA}.  In this formulation, ADHM data defining an instanton of charge $I$ is a linear map $A(z):W\rightarrow V$ depending linearly on $z\in \mathbf{C}^4$, where $W$ is an $I$-dimensional complex vector space with an antilinear map $\sigma_W$ satisfying $\sigma_W^2=1$, and $V$ is a $(2I+2)$-dimensional complex vector space with another antilinear map $\sigma_V$ satisfying $\sigma_V^2=-1$.  If we identify $\mathbf{C}^4$ with $\mathbf{H}^2$ in the standard way, $(z_1,z_2,z_3,z_4)\rightarrow (z_1+jz_2,z_3+jz_4)=(x,y)$, then we define a map $\sigma$ on $\mathbf{C}^4$ to be right multiplication by $j$.  To define an instanton, $A$ must then satisfy:
\begin{enumerate}
\item $A(\sigma z)(\sigma_W w)=\sigma_V(A(z)w)$, 
\item $A(z)$ is injective and $A(z)W$ is an isotropic subspace of $V$ for all $z\in \mathbf{C}^4\setminus \{ 0\}$.  This means that $\sigma_V v=0$ for all $v\in A(z)W$.
\end{enumerate}
There is a prescription for turning this into the standard quaternionic matrix form of ADHM data (see for example \cite{Atiyah}, \cite{Ward}).  First, we can write $W$ as $W_\mathbf{R}\otimes_\mathbf{R}\mathbf{C}$, where $W_\mathbf{R}$ is the subspace of $W$ left fixed by $\sigma_W$.  We can also identify $V$ with an $(I+1)$-dimensional right quaternionic vector space $V_\mathbf{H}$, with multiplication by $j$ given by $\sigma_V$.  Condition 1 above means that we can consider $A$ to be a quaternionic linear map $\mathbf{H^2}\otimes_\mathbf{R}W_\mathbf{R}\rightarrow V_{\mathbf{H}}$. To write out the corresponding matrix, we first choose a basis for the vector space $W_\mathbf{R}$ (a so-called `real basis'), and a basis for $V_\mathbf{H}$ which is unitary with the respect to the standard quaternion inner product.  Now with respect to these bases, define a matrix $C$ whose columns are the images under $A$ of $(1,0)\otimes$ basis vectors of $W_\mathbf{R}$, so $C$ is an $(I+1)\times I$ quaternionic matrix.  Similarly define a matrix $D$ whose columns are the images under $A$ of $(0,1)\otimes$ basis vectors of $W_\mathbf{R}$.  Then $A$ is described by a quaternionic matrix function of the coordinates $x, y$ on $\mathbf{H}^2$:
\[
A(x,y)=Cx+Dy
\]
Condition 2 above is equivalent to requiring that $A(x,y)^\dagger A(x,y)$ be a real non-singular matrix for all $x,y$.  We can obtain standard form ADHM data by setting $y=1$ and finding $R\in Sp(I+1,\mathbf{H})$ and $S\in GL(I,\mathbf{R})$ such that
\[
RCS=\begin{pmatrix}
0 \\
1_I
\end{pmatrix}.
\]
Then the standard quaternionic matrix defining an instanton is $RDS$.

We'll illustrate the `swap' at the ADHM level by writing $p=3/2$ Braam-Austin data in standard quaternionic matrix form. Following \cite{BA}, if the hyperbolic monopole has charge $N$ then in the $p=3/2$ case $W$ has complex dimension $3N$ and $V$ has complex dimension $6N+2$.  Under the circle action, $W$ breaks up into subspaces $W_{-1}\oplus W_0\oplus W_1$ where each $W_i$ has dimension $N$, while $V$ breaks up into subspaces $V_{-3}\oplus V_{-1}\oplus V_{1}\oplus V_{3}$, where $V_3, V_{-3}$ are $N+1$-dimensional and $V_1, V_{-1}$ are $2N$-dimensional.  The subscripts here refer to the weights of the circle action on these spaces.  We can write $A(z)=\sum_{i=1}^4A_iz_i$, so the $A_i$ are matrix components of the map $A$.  One can choose bases for $W$ and $V$ such that:
\[
A_1=\bordermatrix{
~ & W_{-1} & W_0 & W_1 \cr
V_{-3} & I_k & & \cr
 & 0 & & \cr
V_{-1} & & I_k & \cr
 & & 0_k & \cr
V_1 & & & 0_k \cr
 & & & I_k \cr
V_3 & & & 0_k \cr
 & & & 0
}
\]
\[
A_3=\begin{pmatrix}
\beta_{-2} & & \\
v & & \\
 & \beta_0 & \\
 & \gamma_{-1} & \\
 & & -\gamma_{1} \\
 & & \beta_2 \\
 & & 0_k \\
 & & 0
\end{pmatrix}
\]
where $\beta_i,\gamma_i$ are complex $k\times k$ matrices and $v$ is a complex $k$-row vector.  These matrices will correspond to the quaternionic matrices $C,D$ described above, after we make our quaternionic identifications.  

First we need to identify a real basis for $W$.  Suppose that $A_1, A_3$ above are defined with respect to bases $\{ \mathbf{e}^{j}_\alpha\}_{1\leq \alpha\leq k}$ for each $W_j$.  These bases are chosen in such a way that the real structure $\sigma_W: W_j\to W_{-j}$ is just conjugation.  With respect to this real structure, a real basis for $W_{-1}\oplus W_1$ is $\{ \mathbf{e}^{-1}_\alpha\oplus \mathbf{e}^{1}_\alpha\}_{1\leq \alpha\leq k}\cup \{ i\mathbf{e}^{-1}_\alpha\oplus -i\mathbf{e}^{1}_\alpha\}_{1\leq \alpha\leq k}$, while $\{ \mathbf{e}^{0}_\alpha\}_{1\leq \alpha\leq k}$ is already a real basis for $W_0$.  The prescription above says that we need to find the images of these vectors in $V$ and then use the antilinear map on $V$ to identify these images with quaternionic row vectors.  The basis of $V$ used to define $A_1,A_3$ above is chosen such that $\sigma_V$ acts on $V_{-j}\oplus V_{j}$ by $(\mathbf{w},\mathbf{v})\rightarrow (-\overline{\mathbf{v}},\overline{\mathbf{w}})$.  This means we should identify $(\mathbf{w},\mathbf{v})\in V_{-j}\oplus V_{j}$ with the quaternionic vector $\mathbf{w}+j\mathbf{v}$.  The map $\sigma_V$ now corresponds to multiplication by $j$ on the right.  Using the matrices $A_1,A_3$ to determine the images of the real basis vectors gives:
\begin{equation}\label{nonstd}
Cx+Dy=\begin{pmatrix}
0 & 0 & 0\\
I_k & 0_k & iI_k\\
0_k & I_k & 0_k\\
jI_k & 0_k & I_k
\end{pmatrix} x+
\begin{pmatrix}
v & 0 & iv\\
\beta_{-2} & 0_k & i\beta_{-2}\\
-j\gamma_1 & \beta_0 & -k\gamma_1\\
j\beta_2 & \gamma_{-1} & k\beta_2
\end{pmatrix}y
\end{equation}
Multiplying on the left by
\begin{equation}\label{R}
\begin{pmatrix}
1 & 0 & 0 & 0\\
0 & 1/\sqrt{2} & 0 & -j/\sqrt{2}\\
0 & 0 & 1 & 0\\
0 & -i/\sqrt{2} & 0 & -k/\sqrt{2}
\end{pmatrix}
\end{equation} and on the right by
\begin{equation}\label{S}
-\begin{pmatrix}
1/\sqrt{2} & & \\
 & 1 & \\
 & & 1/\sqrt{2}
\end{pmatrix}
\end{equation}
and setting $y=1$ gives ADHM data in standard form:
\begin{equation}\label{std3}
\Delta(x)=\begin{pmatrix}
-v/\sqrt{2} & 0 & -iv/\sqrt{2}\\
-(\beta_{-2}+\beta_2)/2 & j\gamma_{-1}/\sqrt{2} & -i(\beta_{-2}-\beta_2)/2\\
j\gamma_1/\sqrt{2} & -\beta_0 & k\gamma_1/\sqrt{2}\\
i(\beta_2-\beta_{-2}) & k\gamma_{-1}/\sqrt{2} & -(\beta_{-2}+\beta_2)/2
\end{pmatrix}-\begin{pmatrix}
0 & 0 & 0\\
1 & & \\
 & 1 & \\
 & & 1
\end{pmatrix}x
\end{equation}
Imposing the usual ADHM constraint that $\Delta(x)^\dagger\Delta(x)$ be a real non-singular matrix gives exactly the Braam-Austin equations for the $p=3/2$ system.  

Now we specialise to the case of a charge 1 $p=3/2$ monopole.  As described in section \ref{BAconstr}, we can choose our gauge such that $\gamma_i=\gamma\in \mathbf{R}^+,$ $\beta_i=\beta,$ $v=q\gamma$ are constant for all $i$.  We can then interpret $(\beta,\gamma)$ as centre of mass coordinates for the 1-monopole in the half-space model.  We choose our monopole to sit at $\beta=0,\gamma=1$, and we also choose $q=-j$.  Then \eqref{std3} gives us ADHM data in standard form:
\begin{equation}\label{c3} \frac{1}{\sqrt{2}}\begin{pmatrix}
j & 0 & -k\\
0 & j & 0\\
j & 0 & k\\
0 & k & 0
\end{pmatrix}
\end{equation}
Note that the resulting ADHM matrix only has non-zero $j,k$-parts, which shows that this data is invariant under rotations in the $x_1x_2$ plane.  This example illustrates how to interpret the `swap' at the level of ADHM data.  We arranged the Braam-Austin data for a 1-monopole of mass 3/2, which is a set of matrices defined on a lattice with 3 sites, into a single $4\times 3$ matrix with pure $j,k$ entries.  The Braam-Austin data for a $p=1/2$ monopole just consists of a single complex matrix satisfying the ADHM constraints, which, up to an overall factor of $j$, is what we obtained in \eqref{c3}.  This whole construction generalises straightforwardly to give the data \eqref{gen}.

The construction also gives the matrices compensating for axial symmetry.  One can see from \eqref{std3} that the rotation $x\rightarrow e^{i\theta/2}xe^{-i\theta/2}$ acts on the Braam-Austin data as: 
\begin{equation}\label{axsym1}
\beta_j\rightarrow \beta_j,\gamma_j\rightarrow                e^{i\theta}\gamma_j,v\rightarrow v
\end{equation}
We can compensate for this by the gauge transformation
\begin{equation}\label{gtrans}
\begin{aligned}
g_2&=e^{-i\theta}\\
g_0&=1\\
g_{-2}&=e^{i\theta}\\
q&=e^{i\theta}
\end{aligned}
\end{equation}
We need to understand how to interpret gauge transformations in terms of compensating matrices for standard ADHM data.  Gauge transformations act on both $V$ and $W$; in terms of the bases chosen above, the gauge transformation acts on $V$ by:
\begin{equation}\label{Vcomp}
\bordermatrix{
~ & V_{-3} & & V_{-1} & & V_1 & & V_3 & \cr
V_{-3} & g_{-2} &  & & & & & & \cr
 & & q & & & & & & \cr
V_{-1} & & & g_0 & & & & & \cr
 & & & & g_{-2} & & & & \cr
V_1 & & & & & g_0 & & & \cr
 & & & & & & g_2 & & \cr
V_3 & & & & & & & g_2 & \cr
 & & & & & & & & q  
}
\end{equation}
and on $W$ by:
\begin{equation}\label{Wcomp}
\bordermatrix{
~ & W_{-1} & W_0 & W_1 \cr
W_{-1} & g_{-2}^{-1} & & \cr
W_{0} & & g_0^{-1} & \cr
W_1 & & & g_2^{-1} \cr
}
\end{equation}
Using our quaternionic identifications above, \eqref{Vcomp} becomes:
\[
\begin{pmatrix}
q & & & \\
& g_{-2}  & & \\
 & & g_0 & \\
 & & & g_{-2}
\end{pmatrix}
\]
while \eqref{Wcomp} becomes:
\[
\begin{pmatrix}
\cos\theta & 0 & \sin\theta \\
0 & 1 & 0\\
-\sin\theta & 0 & \cos\theta
\end{pmatrix}
\]
and one can check that these matrices will compensate for the transformations \eqref{axsym1} applied to the data \eqref{nonstd}.  Using our transformations \eqref{R} and \eqref{S} gives the standard compensating matrices for the axial symmetry:
\begin{multline}
\begin{pmatrix}
e^{-2i\theta} &  &  & \\
 & \cos\theta & 0 & -\sin\theta \\
 & 0 & 1 & 0\\
 & \sin\theta & 0 & \cos\theta
\end{pmatrix}\frac{1}{\sqrt{2}}\begin{pmatrix}
j & 0 & -k\\
0 & j & 0\\
j & 0 & k\\
0 & k & 0
\end{pmatrix}\begin{pmatrix}
\cos\theta & 0 & \sin\theta \\
0 & 1 & 0\\
-\sin\theta & 0 & \cos\theta
\end{pmatrix}=
e^{i\theta}\frac{1}{\sqrt{2}}\begin{pmatrix}
j & 0 & -k\\
0 & j & 0\\
j & 0 & k\\
0 & k & 0
\end{pmatrix}
\end{multline}
and this construction generalises straightforwardly to give the compensating matrices \eqref{rep}.
\section*{Acknowledgements}
Many thanks to Stefano Bolognesi, Paul Sutcliffe, and my supervisor Richard Ward for useful discussions.  This work was financially supported by an STFC PhD studentship.

\end{document}